\documentclass[acmsmall,screen]{acmart}

\AtBeginDocument{%
  }

\usepackage{amsmath,amsfonts}
\usepackage{graphicx}
\usepackage{textcomp}
\usepackage{xcolor}
\usepackage{caption}
\usepackage{lipsum}
\usepackage{makecell}
\usepackage{xspace}
\usepackage{float}
\usepackage{tipa}

\usepackage{arydshln}
\usepackage{enumitem}
\usepackage{multirow}
\usepackage{tcolorbox}
\usepackage{colortbl}
\usepackage[export]{adjustbox} 
\usepackage[T1]{fontenc}
\usepackage{tabularx}
\usepackage{threeparttable}
\usepackage{listings}
\usepackage{xcolor}
\usepackage{booktabs}
\usepackage{subcaption}
\usepackage{algorithm}
\usepackage{algorithmic}

\usepackage{changes}
\newcolumntype{Y}{>{\raggedright\arraybackslash}X}

\newcommand{\methodname}{TDDev\xspace}
\newcommand{\taskname}{Req-to-App\xspace}
\definecolor{codegreen}{rgb}{0,0.6,0}

\newsavebox{\arrangebox}

 \newcommand{\advColor}[1]{%
  \pgfmathsetmacro{\val}{#1}%
  \ifdim \val pt < 0pt
    \textcolor{red}{#1}%
  \else
    \ifdim \val pt > 5pt
      \underline{\textcolor{codegreen}{#1}}%
    \else
      \textcolor{codegreen}{#1}%
    \fi
  \fi
}
\setcopyright{acmlicensed}
\copyrightyear{2018}
\acmYear{2018}
\acmDOI{XXXXXXX.XXXXXXX}
\acmConference[Conference acronym 'XX]{Make sure to enter the correct
  conference title from your rights confirmation email}{June 03--05,
  2018}{Woodstock, NY}

\begin{document}

\title{Automatically Generating Web Applications from Requirements Via Multi-Agent Test-Driven Development}

\author{Yuxuan Wan}
\authornote{Both authors contributed equally to this research.}
\affiliation{%
  \institution{The Chinese University of Hong Kong}
  \city{Hong Kong}
  \country{China}
}
\email{yxwan9@cse.cuhk.edu.hk}

\author{Tingshuo Liang}
\authornotemark[1]
\affiliation{%
  \institution{The Chinese University of Hong Kong}
  \city{Hong Kong}
  \country{China}
}
\email{1155210994@link.cuhk.edu.hk}

\author{Jiakai Xu}
\affiliation{%
  \institution{Columbia University in the City of New York}
  \city{New York}
  \country{USA}
}
\email{ax2155@columbia.edu}

\author{Jingyu Xiao}
\affiliation{%
  \institution{The Chinese University of Hong Kong}
  \city{Hong Kong}
  \country{China}
}
\email{whalexiao99@gmail.com}

\author{Yintong Huo}
\authornote{Yintong Huo is the corresponding author.}
\affiliation{%
  \institution{Singapore Management University}
  \city{Singapore}
  \country{Singapore}
}
\email{ythuo@smu.edu.sg}

\author{Michael Lyu}
\affiliation{%
  \institution{The Chinese University of Hong Kong}
  \city{Hong Kong}
  \country{China}
}
\email{lyu@cse.cuhk.edu.hk}


\begin{abstract}
Developing full-stack web applications is complex and time-intensive, demanding proficiency across diverse technologies and frameworks. Although recent advances in multimodal large language models (MLLMs) enable automated webpage generation from visual inputs, current solutions remain limited to front-end tasks and fail to deliver fully functional applications. In this work, we introduce \methodname, the first test-driven development (TDD)-enabled LLM-agent framework for end-to-end full-stack web application generation. Given a natural language description or design image, \methodname automatically derives executable test cases, generates front-end and back-end code, simulates user interactions, and iteratively refines the implementation until all requirements are satisfied. Our framework addresses key challenges in full-stack automation, including underspecified user requirements, complex interdependencies among multiple files, and the need for both functional correctness and visual fidelity. Through extensive experiments on diverse application scenarios, \methodname achieves a 14.4\% improvement on overall accuracy compared to state-of-the-art baselines, demonstrating its effectiveness in producing reliable, high-quality web applications without requiring manual intervention. The code of \methodname is available at \url{https://github.com/yxwan123/TDDev}.
\end{abstract}



\maketitle

\section{Introduction}

In the modern digital era, web applications play a pivotal role as foundational platforms that support a wide range of everyday activities. The scale of this ecosystem is immense: recent reports estimate more than 1.1 billion active websites, with an additional 252,000 new sites launched daily~\cite{website_statistics_2024, wordpress_statistics_2024}.

Developing web applications, also referred to as full-stack development, involves two stages: front-end development and back-end development. The former is concerned with the graphical user interface (GUI), such as layouts, content, and interactive elements~\cite{Metwalli2025FrontEnd}. The latter focuses on the server-side logic that supports the application, including APIs, data storage, and request handling~\cite{AWSFrontVsBack, AWSFullStack}.

Full-stack development is complicated and time-consuming. It requires proficiency in both client-side and server-side technologies and constant adaptation to evolving tools and frameworks~\cite{Simplilearn2025FullStack}. For novices, the breadth of skills needed, e.g., HTML, CSS, JavaScript, databases, and server-side languages, creates a steep learning curve that hinders turning ideas into applications~\cite{Simplilearn2025FullStack}. Even for experienced developers, the sheer quantity of frameworks, programming languages, and deployment models leads to cognitive overload~\cite{Chen2018FromUI, Nguyen2015ReverseEM, Lelli2015ClassifyingAQ, Moran2018AutomatedRO}; industry data show that learning a new development environment can take as long as a year~\cite{ITRevolution2021FullStack}. These challenges necessitate an automated solution framed as \textit{\textbf{\taskname}: automatically translating user requirements (e.g., text descriptions, design sketches) into a functional web application}. 

\begin{table}[t]
\centering
\small
\begin{adjustbox}{max width=\linewidth}
\begin{tabular}{@{}p{2cm}p{6.5cm}p{6.5cm}@{}}
\toprule
\textbf{Aspect} & \textbf{Front-End Development} & \textbf{Full-Stack Development} \\
\midrule
\textbf{Goal}  & Show information and visuals in the browser & Front-End + read/write data and enforce rules \\
\textbf{Output} & Static webpage (a single HTML file) & Functional application (a project with multiple files) \\
\textbf{Platform} & Browser only & Browser + server \\
\textbf{Data} & No persistent data (or mock data embedded) & Real data stored \\
\textbf{Examples} & Marketing landing page; Personal homepage & Booking system; forum with posts/replies \\
\textbf{Related Work} & \cite{Chen2018FromUI, Nguyen2015ReverseEM, Abdelhamid2020DeepLP, Moran2018MachineLP, zhou2024bridging, beltramelli2018pix2code, Si2024Design2CodeHF, Gui2024VISION2UIAR, Zhou2024BridgingDA, Wu2025MLLMBasedUA, Wan2025DivideAC, Wan2024MRWebAE, xiao2024interaction2code, Xiao2025DesignBenchAC, jiang2025screencoder} & \cite{lu2025webgen}(benchmark study), \textbf{Ours}  \\

\bottomrule
\end{tabular}
\end{adjustbox}
\caption{Differences between front-end and full-stack development.}
\label{tab:fron-vs-full}
\end{table}

Despite its importance, practical solutions for building full-stack web applications from user requirements remain underexplored. 

The closest line of work focuses on the simpler “design-to-code” task, which only produces front-end webpages rather than full-fledged applications. Table~\ref{tab:fron-vs-full} compares front-end and full-stack development and highlights related research. Within this scope, prior studies have focused on converting GUI designs into code for Android apps~\cite{Chen2018FromUI, Nguyen2015ReverseEM, Abdelhamid2020DeepLP, Moran2018MachineLP, zhou2024bridging}, or on generating synthetic datasets and simple designs~\cite{beltramelli2018pix2code} to train deep learning models (e.g., CNN, LSTM). More recently, Multimodal Large Language Models (MLLMs) have shown promise in generating webpage code from screenshots~\cite{Si2024Design2CodeHF, Zhou2024BridgingDA, Wu2025MLLMBasedUA, Wan2025DivideAC, Wan2024MRWebAE, Gui2025UICopilotAU}. However, full-stack development introduces additional challenges beyond layout rendering, including implementing backend logic, managing servers and databases, handling APIs and resources, and ensuring seamless integration across multiple components. Existing design-to-code methods thus fall short in addressing these requirements, leaving the challenges of full-stack automation largely unexplored.


Apart from research works, commercial tools such as Bolt.new\footnote{https://bolt.new}
 and Lovable.dev\footnote{https://lovable.dev}
 are able to generate complete websites from user input. Yet, their performance remains limited: a recent empirical study~\cite{lu2025webgen} reports that applications generated by these frameworks with state-of-the-art LLMs fail to implement required functionalities or even fail to compile in over 70\% of cases. 
 As a result, users must manually devise test cases, check functionality, describe errors or visual mismatches through the chat interface, and repeatedly request refinements until the application meets expectations—a process that is slow, labor-intensive, and frustrating~\cite{Becker2025METR}.
 
\begin{figure}
    \centering
\includegraphics[width=0.9\linewidth]{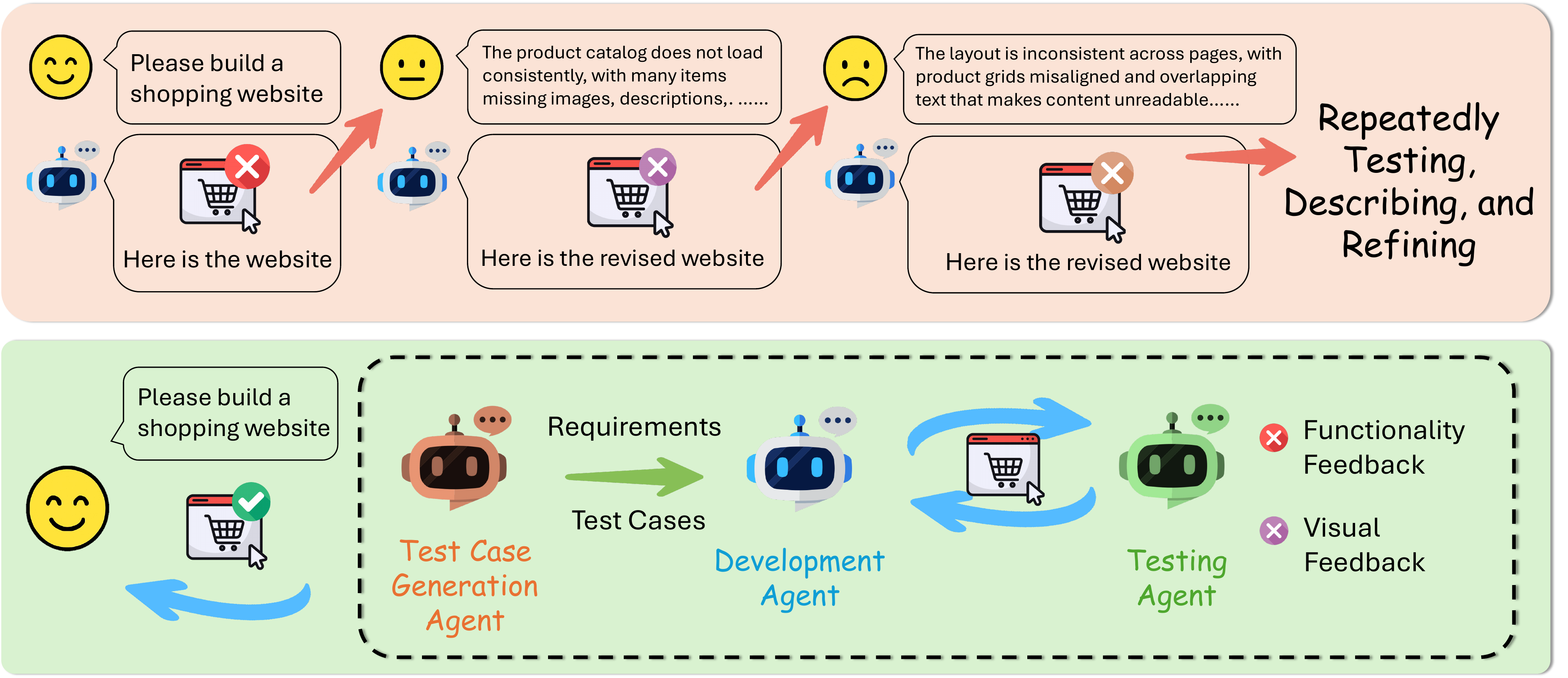}
    \caption{Comparison between our proposed \methodname framework (lower) and current industry tools (upper).}
    \label{fig:ttd-advantage}
\end{figure}

Test-driven development (TDD) is a software engineering practice where developers iteratively write a test for a specific feature, implement code to satisfy that test, and then refine the code to improve quality~\cite{Mathews2025Test}. Inspired by TDD, we believe 
that an agent-based system can collaboratively emulate such a workflow to handle \taskname without further user intervention.
Specifically, we expect the development agents to: \textbf{1) generate test cases from user requirements to validate correctness and functionality, 2) produce full-stack code and test it against the generated tests, and 3) iteratively refine the code based on test results until all cases pass and the application meets the desired standards.} Figure~\ref{fig:ttd-advantage} shows a comparison between our proposed pipeline and current industry tools.

Unfortunately, creating a TDD framework in the context of web development presents several unique challenges. \textbf{First}, test case generation methods in traditional TDD are not directly applicable to the \taskname scenario because (1) user requirements are usually high-level and under-specified (e.g., “Please create a shopping website”), making it difficult to derive comprehensive, executable test cases (e.g., user registration, shopping cart utilities); (2) web testing focuses more on non-crash defects such as interaction and logical errors, which often evade conventional TDD techniques~\cite{Liu2024VisiondrivenAM}. This requires test cases to capture user interaction flows (e.g., entering a username and password, clicking login) rather than simple input–output pairs. \textbf{Second}, web application test cases must simultaneously account for user interaction flows and UI layout checks, which traditional script-based or unit-test TDD strategies cannot adequately capture~\cite{Yu2023VisionBasedMA}. Addressing these requirements calls for end-to-end agents capable of performing fine-grained element recognition, identifying usability issues that are technically functional but practically unusable (e.g., low contrast, overlapping elements), and remaining robust to unexpected scenarios such as \textit{login walls}—barriers that require authentication before granting access to subsequent functionality. \textbf{Finally}, beyond testing, the framework must deliver rich feedback covering both functional correctness and visual presentation to effectively guide refinement—an aspect that has also been unexplored in existing work. Table~\ref{tab:trad-vs-web} provides a comprehensive comparison between TDD on traditional code tasks and TDD in \taskname scenario.

In this work, we introduce \textbf{\methodname, the first TDD-enabled LLM-agent framework for full-stack web development}, which automatically generates high-quality web applications from natural language descriptions or image designs without requiring further user intervention. Specifically, we first design a test case generation agent that leverages requirement engineering and soap-opera test techniques to derive detailed requirements and test cases from high-level user requirements. Then, we implement a development agent that integrates file handling, planning, and memory capabilities for full-stack development. Finally, we create a testing agent capable of simulating user interactions based on provided test cases. Together, these agents are orchestrated in a TDD workflow that generates high-quality web applications from user requirements in an end-to-end manner. 

To evaluate the effectiveness of \methodname, we propose a dataset \textbf{\taskname-MM}, which is an augmented version of WebGen-Bench~\cite{lu2025webgen}, a benchmark designed to assess LLM agents’ ability to generate multi-file website codebases from text requirements. We extended WebGen-Bench to a multimodal setting—where users can provide both design images and text for higher-fidelity requirements. We evaluated \methodname on the extended benchmark, and the results show that  \methodname achieves a 14.4\% improvement on overall accuracy compared to state-of-the-art baselines, demonstrating its effectiveness in producing reliable, high-quality web applications without requiring manual intervention. 

In summary, our contributions are as follows:
\begin{itemize}
    \item We introduce \methodname, the first TDD-enabled LLM-agent framework for full-stack web development, addressing the unique challenges of integrating TDD into web application generation.
    \item We design and implement an orchestration of three specialized agents: (1) a test case generation agent that transforms under-specified user requirements into executable test cases, (2) a development agent with planning, memory, and file-handling capabilities for multi-file full-stack projects, and (3) a testing agent that simulates realistic user interactions and provides feedback on both functionality and visual presentation.
    \item We conduct extensive experiments across diverse web development scenarios, showing that \methodname substantially improves generation quality, achieving a 14.4\% higher overall accuracy, and significantly decreases human development effort in the user study.
    \item We release the implementation of \methodname, to foster future research on TDD-driven web application generation and support practical development tasks.
\end{itemize}

\section{Background}

\subsection{Task Formulation: \taskname}
 This task takes a piece of text description and optionally a visual webpage design as input, aiming to generate code that satisfies the description. Let $T_0, I_0$ represent the text description and image (optional) input by the user. Given $T_0, I_0$, an agent generates a web application $App = Agent(I_0, T_0)$. The functionality of $App$ should closely match $T_0$, and the GUI of $App$ should resemble  $I_0$.

\subsection{Related Work}
\subsubsection{UI Code Generation}
UI code generation produces front-end code from screenshots or design images. Early approaches typically rely on CNNs and Computer Vision (CV) techniques for automated GUI prototyping~\cite{acsirouglu2019automatic, Cizotto2023WebPF, Moran2018MachineLP, Xu2021Image2e, Chen2018FromUI, nguyen2015reverse, beltramelli2018pix2code, Chen2022CodeGF}. The advent of MLLMs has enabled more advanced approaches, such as Design2Code~\cite{Si2024Design2CodeHF}. To address element omission, distortion, and misarrangement, DCGen~\cite{Wan2025DivideAC} introduced a divide-and-conquer strategy, while LayoutCoder~\cite{Wu2025MLLMBasedUA} and UICopilot~\cite{Gui2025UICopilotAU} adopted layout-aware techniques. DeclarUI~\cite{zhou2024bridging} further incorporated page transition graphs for mobile UI generation. ScreenCoder~\cite{jiang2025screencoder} implement a modular agent system and reinforcement learning (RL) framework for front-end development. Interaction2Code and DesignBench~\cite{Xiao2024Interaction2CodeHF, Xiao2025DesignBenchAC} added interaction-aware generation and repair, and MRWeb~\cite{Wan2024MRWebAE} explored resource-aware generation. EfficentUICoder~\cite{xiao2025efficientuicoder} proposed a token compression framework for efficient UI code generation. Despite these advances, existing methods remain limited to front-end webpages without functionality. Recently, an empirical study, WebGenBench~\cite{lu2025webgen}, benchmarked industry tools on full-stack development, revealing that applications generated with state-of-the-art LLMs fail to implement required functionalities—or even compile—in over 60-70\% of cases, underscoring the need for more effective approaches to improve the performance of LLM-based agents in automatic full-stack web development.

\subsubsection{LLM Agents}
The rise of Large Language Model (LLM) agents~\cite{fang2025cognitive, xia2025demystifying} has opened new opportunities for automation in software engineering. These agents have demonstrated effectiveness in code generation~\cite{dong2025survey}, program repair~\cite{bouzenia2024repairagent}, and program testing~\cite{feldt2023towards}. Building on LLM agents, commercial tools such as Bolt.new\footnote{https://bolt.new}
 and Lovable.dev\footnote{https://lovable.dev}
 are able to generate complete websites from user input. However, their performance is limited and requires extensive user intervention, which is frustrating~\cite{lu2025webgen, Becker2025METR}. Figure~\ref{fig:ttd-advantage} shows a comparison between our proposed pipeline and current industry tools.

\subsubsection{GUI Testing}
Automated GUI testing has been studied through multiple approaches. Record-and-replay methods are straightforward to implement but tend to be fragile and require frequent maintenance as applications evolve~\cite{Yu2023VisionBasedMA}. Random-based tools such as Monkey~\cite{android_monkey} reduce manual effort but often achieve limited functional coverage. Model-based testing~\cite{Miguel2016GUIAU, Gu2019PracticalGT} introduces more structure by deriving cases from formal models, though its effectiveness is restricted by limited model accuracy and continuous model updates, and it generally overlooks GUI semantics. Learning-based methods~\cite{Lan2024DeeplyRA, Pan2020ReinforcementLB, Li2019HumanoidAD}, often built on reinforcement learning, can discover testing policies but require substantial training data and adapt less effectively to rapidly changing applications with a lack of deeper semantic understanding~\cite{Liu2023MakeLA}. Recently, MLLM-based methods~\cite{Liu2023MakeLA, Liu2024VisiondrivenAM} have started to leverage visual semantics and functional structures, offering a promising direction for GUI testing. Their planning, however, remains largely exploratory, making it difficult to achieve systematic coverage of targeted functionalities (e.g., evaluating the usability of a shopping cart). Moreover, most prior work has centered on the Android platform, whereas our focus is on general web applications, where testing must address both overall UI layout and fine-grained component behavior within the broader \taskname task, rather than the more tailored, touch-oriented layouts of mobile interfaces.

\subsubsection{Test Driven Development}
Improving code generation with tests has become a popular direction in traditional code tasks. Wang et al.~\cite{Wang2022TestDriven} use test execution feedback during training to detect errors, while Chen et al.’s Codex~\cite{Chen2021Evaluating} highlights the limits of single-sample generation and motivates iterative testing. Compiler feedback has been exploited to ensure compilable code~\cite{Wang2022Compilable}. AutoCodeRover~\cite{Zhang2024AutoCodeRover} applies test cases for spectrum-based fault localization. Mathews et al.~\cite{Mathews2025Test} empirically demonstrated that applying TDD principles to LLM-based code generation is beneficial using human-written tests. Liu et. al~\cite{liu2024write} designed an automated test-driven checker development approach with LLM. However, all of these works focus on traditional code tasks, and most of them on program repair. Due to the fundamental difference between web development and traditional code generation, as shown in Table~\ref{tab:trad-vs-web}, none of these methods applies to the full-stack development scenario.

\begin{table}[h]
\centering
\small
\caption{Comparison between traditional TDD and web application TDD.}
\label{tab:trad-vs-web}
\begin{adjustbox}{max width=\textwidth}
\begin{tabular}{@{}p{2cm}p{6.5cm}p{6.5cm}@{}}
\toprule
\textbf{Aspect} & \textbf{Traditional Code TDD} & \textbf{Web Application TDD} \\ \midrule
\textbf{Scope} & Single function, class, or algorithm in isolation. & Integration of front-end and back-end. \\
\textbf{Test Cases} & Clear unit tests with well-defined input–output pairs. & Derived from vague user requirements and expanded into functional and visual validations.  \\
\textbf{Test Methods} & Automated unit tests or compiler checks. & End-to-end user simulation and UI visual evaluation. \\
\textbf{Artifacts} & Single source file or self-contained program. & Multiple interdependent files (HTML, CSS, JavaScript, server code, database, configuration). \\
\textbf{Feedback} & Binary results (pass/fail) or compilation errors. & Rich feedback on both functional correctness and visual presentation.\\
\textbf{Example} & Sorting an array, computing shortest path, or matrix multiplication. & Shopping website with product catalog, cart, login integration, and responsive UI. \\
\textbf{Related Work} & \cite{Wang2022TestDriven, Chen2021Evaluating, Wang2022Compilable, Zhang2024AutoCodeRover, Mathews2025Test, liu2024write} & \textbf{Ours}
\\ \bottomrule
\end{tabular}
\end{adjustbox}
\end{table}

\section{Methodology}
This section introduces the mechanism of \methodname, a framework that automatically generates high-quality web applications from natural language descriptions or image designs. \methodname begins with a test case generation agent, which applies requirement engineering and soap-opera testing techniques to derive detailed requirements and test cases from high-level user inputs. It then employs a development agent equipped with file handling, planning, and memory capabilities to perform full-stack development. Finally, a testing agent inspects the generated application by simulating user interactions according to the test cases. These agents are orchestrated within a TDD workflow that iteratively refines the application based on testing feedback until the output meets the desired standards.

\begin{figure}
    \centering
    \includegraphics[width=0.9\linewidth]{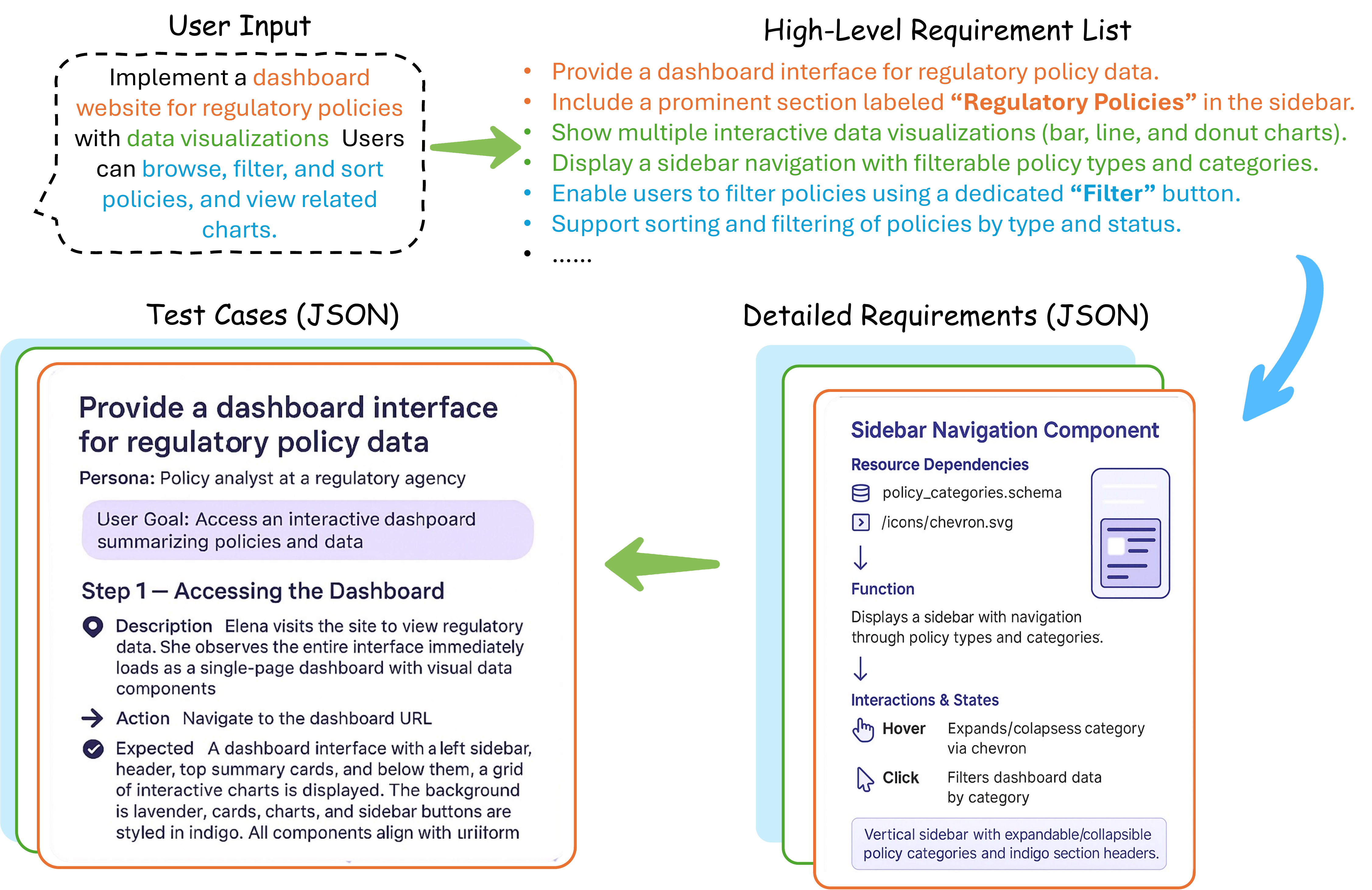}
    \caption{The workflow of the test generation agent.}
    \label{fig:test-gen-agent}
\end{figure}

\subsection{Test Generation Agent}
\label{sec:test-generation-agent}
 This stage aims to refine the input into a set of actionable and concrete requirements, each paired with a corresponding test case. Figure~\ref{fig:test-gen-agent} illustrates the workflow. The agent takes as input a high-level web development instruction, optionally accompanied by a screenshot of the expected final web application. It then progressively transforms it into concrete specifications ready for testing. 

The workflow is orchestrated across three specialized agents. The \textbf{requirement decomposition agent} first parses the input and produces a structured list of discrete requirements. Each requirement is expressed as a concise statement, typically describing a functionality, layout constraint, or design element. Beyond explicit instructions, the agent also infers implicit requirements necessary to make the application complete and feasible.  

Next, the \textbf{requirement elaboration agent} enriches each high-level item with detailed specifications, covering functionalities, static UI design, and dynamic interactions. At this stage, relevant data sources are also identified, which may include predefined datasets, databases, or external APIs. Depending on the data type, the agent either provides direct content or generates database setup instructions with defined schemas. The refined outputs are represented in JSON format, ensuring determinism and executability in subsequent implementation.  

Finally, the \textbf{test case generation agent} produces one test case for each requirement, ensuring complete coverage and alignment with user expectations. Inspired by the concept of \textit{soap opera testing}~\cite{Kaner2013AnIT}, the agent begins by imagining a user persona with specific goals and then generates step-by-step instructions describing user actions and expected outcomes. These structured cases not only capture realistic usage scenarios but also reduce ambiguity and minimize testing errors.  

Overall, this process simulates industrial development workflows by systematically bridging high-level user requirements and concrete testable specifications, thereby ensuring strong alignment between the final application and the original intent~\cite{Cheng2007ResearchDI}.

\begin{figure}
  \includegraphics[width=0.75\linewidth]{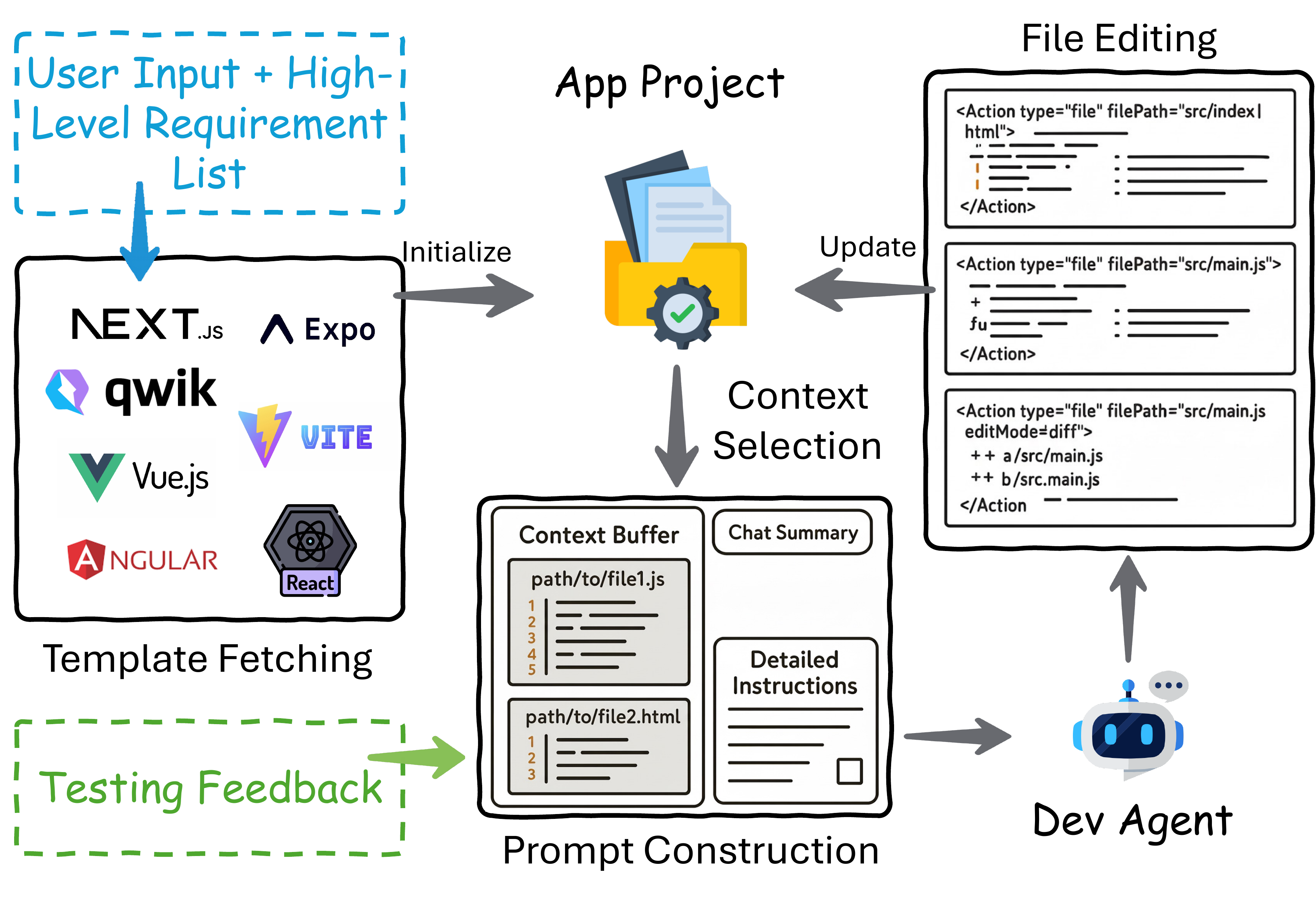}
  \caption{Workflow of the development agent.}
  \label{fig:dev-agent}
\end{figure}

\begin{table}[t]
\centering
\caption{13 different starter templates from GitHub covering various frameworks and use cases.}
\label{tab:templates}
\resizebox{0.6\linewidth}{!}{%
\begin{tabular}{ll}
\toprule
\textbf{Template} & \textbf{Description} \\
\midrule
Expo App          & Cross-platform mobile app development \\
Basic Astro       & Static website generation \\
NextJS Shadcn     & Full-stack Next.js with \texttt{shadcn/ui} components \\
Vite Shadcn       & Vite with \texttt{shadcn/ui} components \\
Qwik TypeScript   & Resumable applications \\
Remix TypeScript  & Full-stack web applications \\
Slidev            & Developer presentations using Markdown \\
SvelteKit         & Fast web applications \\
Vanilla Vite      & Minimal JavaScript projects \\
Vite React        & React with TypeScript \\
Vite TypeScript   & Type-safe development \\
Vue.js            & Vue applications \\
Angular           & Angular applications with TypeScript \\
SolidJS           & Lightweight reactive applications \\
\bottomrule
\end{tabular}%
}
\end{table}
\vspace{-15pt}

\subsection{Development Agent}

Forked from \texttt{Bolt.diy}, we implement a development agent equipped with file management capabilities to support end-to-end application development. The agent takes as input both the initial natural language requirement from the user and the structured high-level requirement list produced by the test generation agent. It then initializes a project from a suitable template, iteratively determines which files should be edited, and applies modifications until the application satisfies the requirements. Figure~\ref{fig:dev-agent} illustrates the workflow of development agent.

\paragraph{Template Fetching}
To bootstrap development, the agent leverages 13 open-sourced starter projects from GitHub, covering a wide range of frameworks and use cases (Table~\ref{tab:templates}). During initialization, the agent prompts an LLM to classify the application type based on the input requirements and select the most appropriate template. The chosen template is then cloned into the working directory, serving as the seed project for subsequent development.

\paragraph{Context Selection}
Once initialized, the agent adapts the seed project to the user requirements or feedback. To do so, it constructs a focused context prompt for code generation and file editing. Context construction proceeds in three steps:  
1) File filtering: heuristics specific to each template exclude irrelevant files (e.g., \texttt{node\_modules}, build artifacts, hidden directories).  
2) Context buffer extraction: the system collects the current state, including (i) all available file paths, (ii) already loaded files, and (iii) the chat summary with the user’s latest request.  
3) Relevance selection: the agent invokes a dedicated LLM instance to analyze the context buffer and determine which files are relevant. The agent is instructed to respond in a structured XML format, explicitly marking included and excluded files. An illustrative example of the selection prompt is shown below:

\begin{tcolorbox}
[colback=white,colframe=black!50,title=Prompt For Selecting Relevant Context]
\footnotesize
You are a software engineer working on a project. You have access to the following files: 

[Available Files] 

You have the following code loaded in the context buffer that you can refer to: 

[Context Buffer]
 
Select only the files relevant to the current task. Respond in the following XML format: 

[Response Format]
\end{tcolorbox}

\paragraph{Prompt Construction} The system parses the LLM response to extract included File and excluded File directives, updating the context buffer by adding newly relevant files and removing those no longer needed for the current development task. The updated context buffer is formatted into a structured prompt. Each included file is wrapped in a file tag with the full file content, creating a comprehensive view of the codebase that the LLM can reference when making modifications. The prompt for development is shown below:

\begin{tcolorbox}
[colback=white,colframe=black!50,title=Prompt for Development]
\footnotesize
You are a software engineer working on a project. Below is the artifact containing the context loaded into context buffer for you to have knowledge of and might need changes to fullfill current user request.

<Context Buffer> 

\hspace{0.5cm} <file filePath="path/to/file1.js">  
[File Content With Line Number]  
</file>

\hspace{0.5cm} <file filePath="path/to/file2.html">  
[File Content With Line Number]  
</file> 

</Context Buffer>

Here is the chat history till now: 

[Chat Summary] 

Please make necessary updates to the project based on the instructions: 

[Detailed Instructions \& Response Format]
\end{tcolorbox}

\paragraph{File Editing} After the development prompt is constructed, the LLM generates a structured response containing one or more tags that specify file operations required to fulfill the user request. Each action corresponds to a file creation, modification, or update operation within the project workspace.

For most cases, file modifications follow a full-content replacement strategy. The LLM outputs the complete updated file content within a \texttt{<Action type="file" filePath="path/to/file">} tag. This approach guarantees consistency, eliminates ambiguity about the final file state, and avoids potential merge conflicts, ensuring that the edited file can be safely written into the workspace without relying on prior versions. In addition, the system optionally supports a diff-based editing mode for scenarios where only small, localized changes are required. In this mode, the LLM encloses changes within file modification tags using unified diff syntax, specifying exactly which lines to add, modify, or remove. Unlike the default full-content replacement, diff-based updates are explicitly invoked and optimized for efficiency when the modified portions are small relative to the file size.

Before applying edits, the system cleans formatting artifacts—such as Markdown code fences and escaped HTML entities—for workspace compatibility. It also enforces file-locking by listing protected files in the system prompt, ensuring the LLM skips any modifications that would compromise project integrity.

\subsection{Testing Agent}
The testing agent takes as input the entire project directory produced by the development agent and the test cases generated by the test case generation agent. It executes these test cases on the application and provides structured feedback to guide further refinement.  

\paragraph{Deployment Verification} The testing agent first launches the application using PM2\footnote{https://pm2.keymetrics.io/}, a widely used process manager for Node.js applications, and hosts it on a local browser. During the launch phase, it performs an initial verification by capturing a screenshot to check for errors such as blank screens or crash messages. If a launch failure is detected, all subsequent tests are aborted, and the agent immediately returns feedback containing logs and diagnostic information. When an expected screenshot is provided, the captured and reference screenshots are compared to identify visual discrepancies, which are then included in the feedback.  

\paragraph{User Simulation} To simulate the interaction flows specified in the soap opera test cases (Section~\ref{sec:test-generation-agent}), we integrate Browser-Use~\cite{browser_use2024}, an LLM-powered autonomous browser agent with 69.7k GitHub stars. Browser-Use enables natural language control of a browser, supporting navigation, form filling, data extraction, and multi-step workflows (e.g., “Book a flight from Hong Kong to New York on the United Airlines website”). For efficiency, test cases are executed in parallel: multiple application instances are launched, and each instance is tested independently by a Browser-Use agent. The agent strictly follows the soap opera testing paradigm, enacting the defined user persona, executing step-by-step actions, verifying outcomes against expectations, and logging any deviations.  

\paragraph{Feedback Construction} After executing the test cases, the testing agent processes the Browser-Use logs to produce comprehensive feedback. Each testing report details the failed steps, executed actions, expected versus actual outcomes, error categories, supplementary technical information, and actionable recommendations to guide correction and prevent recurrence. 

\paragraph{Error Handling} In addition, the testing agent is equipped with autonomous error-handling capabilities. It incorporates a bounded retry mechanism to handle transient failures, which prevents infinite execution loops. We observe that the agent exhibits sufficient environmental awareness and problem-solving capabilities. For instance, upon encountering a login modal with ``account does not exist'' error, it can autonomously attempt to register and authenticate before proceeding with the predefined test steps.

\section{Experiment}
\subsection{Research Questions}
We evaluate the effectiveness of \methodname on automating full-stack web development through answering the following research questions (RQs):
\begin{itemize}
    \item \textbf{RQ1:} How effective is \methodname in automating full-stack web development?
    \item \textbf{RQ2:} Does the multi-step design of the test case generation agent improve effectiveness?
    \item \textbf{RQ3:} Can feedback from the testing agent enhance the quality of generated applications?
    \item \textbf{RQ4:} In what ways does \methodname provide advantages for developers over existing open-source tools?
\end{itemize}

\subsection{Baselines}
We evaluate two widely used open-sourced and proprietary industry-level code-agent frameworks as baselines: Bolt.diy\footnote{https://bolt.new/}
 and Cursor\footnote{https://cursor.com/}
. Bolt.diy, the open-source version of Bolt.new, is a browser-based framework for generating and previewing web applications, with 17.7k stars on GitHub. It provides a user interface with a Linux-like WebContainer and prompts the model to select frontend and backend frameworks (e.g., Vite, React, Remix) before importing and extending a template. Cursor, a proprietary AI-assisted integrated development environment (IDE) built on Visual Studio Code, provides features such as code generation and an agent mode for end-to-end task execution. It has become one of the most widely adopted proprietary coding assistants, reportedly reaching 360k active users by 2024~\cite{endo2024cursor}. In our experiments, we employ Cursor's agent mode, manually supplying the WebGen-Bench requirement as the initial prompt using a subset of 10 WebGen-Bench test data. Whenever the agent raises follow-up questions, we either select the default option or allow the agent to decide autonomously, continuing this process until no further queries are made. 

\subsection{Backbone Models}
We evaluate the frameworks on two state-of-the-art (SOTA) general-purpose proprietary: GPT-4.1 and Claude-4-Sonnet. We also discuss the performance of our framework on open-sourced models such as Qwen-2.5-VL and DeepSeek-V3.1 in Section~\ref{sec:discussion}.

\subsection{Experiment Setup}
All experiments are run on an iMac with 10 Core Intel Core i9 processor, 32GB RAM. All LLM models are accessed through the official API services. Temperatures are set to 0 for all models. Max tokens are set to the maximum allowable value for each model.

\subsection{Cost \& Efficiency}
As shown in Table~\ref{tab:cost-comparison}, all methods require roughly four minutes to develop, but differ in testing and cost structures. Bolt.diy and Cursor rely on manual testing, with low per-round cost (0.18 USD) or flat subscription (20 USD/month) but requiring constant user attention and manual intervention. In contrast, \methodname incurs a higher per-round token cost (0.36 USD) but fully automates testing and refinement, eliminating manual effort and allowing developers to focus on other tasks, making it more efficient despite slightly higher computational cost.

\begin{table}[t]
\centering
\small
\caption{Comparison of development and testing costs across different frameworks per round. One round refers to a full cycle of code generation and feedback.}
\label{tab:cost-comparison}
\begin{tabular}{lccc}
\toprule
\textbf{Method} & \textbf{Develop (Token/Time)} & \textbf{Test (Token/Time)} & \textbf{Cost (Claude-4-Sonnet)} \\
\midrule
Cursor       & Unknown / $\sim$4 min   & Manual & 20 USD/month \\
Bolt.diy     & $\sim$10K / $\sim$4 min & Manual & 0.18 USD/round \\
\methodname  & $\sim$10K / $\sim$4 min & $\sim$10K / $\sim$4 min & 0.36 USD/round \\

\bottomrule
\end{tabular}
\end{table}

\subsection{Evaluation}
\subsubsection{Dataset Construction}
\begin{figure}
    \caption{An example data instance. Only 1 out of 7 test cases are shown.}
    \label{fig:data-instance}
    \centering
    \includegraphics[width=\linewidth]{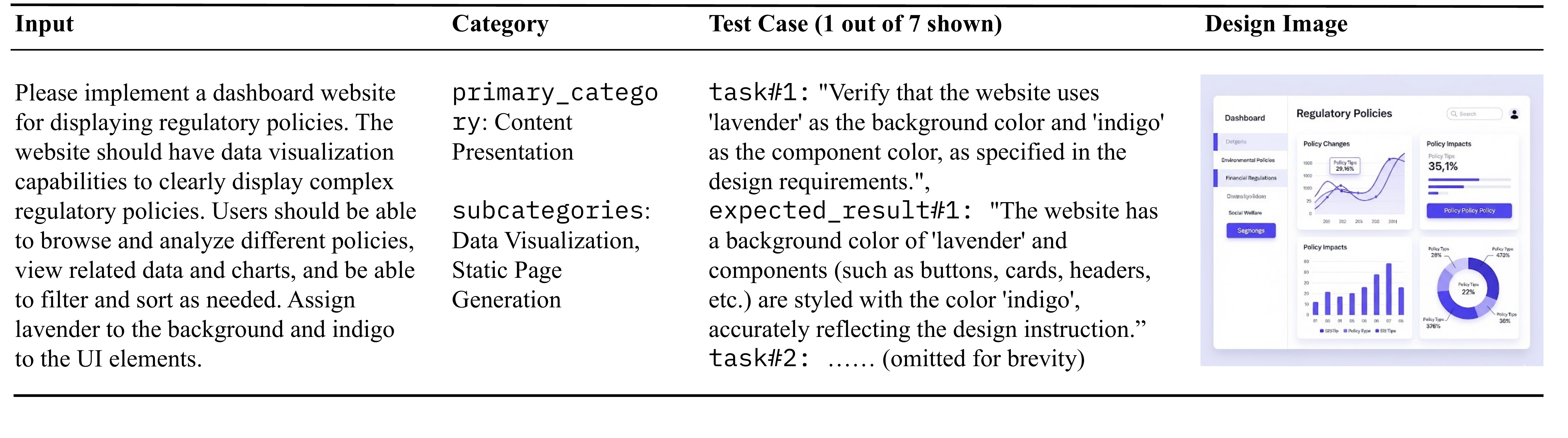}
\end{figure}

We use the test set of WebGen-Bench~\cite{lu2025webgen}, a benchmark designed to assess LLM agents’ ability to generate multi-file website codebases from scratch, as our core evaluation data. However, WebGen-Bench is text-only, with user requirements expressed purely in natural language. To extend it to a multimodal setting—where users can provide both design images and text for higher-fidelity requirements—we augment the benchmark by generating images for each text description using Gemini-2.5-Flash-Image~\cite{google_gemini_api_docs}, a multimodal LLM capable of producing realistic designs from prompts. These images serve as additional visual requirements for the agents. We name this dataset \taskname-MM. During evaluation, we further assess visual consistency between the generated websites and the corresponding design images. Figure~\ref{fig:data-instance} shows an instance of the data, and the statistics of the dataset are shown in Table~\ref{tab:webgen-categories}. To perform the evaluation, we randomly select 10 websites while preserving the category distribution of the test set.

\begin{table}[t]
\centering
\caption{The number of website-generation instructions in each technical category in \taskname-MM. Each main category contains multiple subcategories. A sample may belong to one main category and multiple subcategories.}
\label{tab:webgen-categories}
\resizebox{0.6\linewidth}{!}{
\begin{tabular}{ll|ll}
\toprule
\textbf{Main Categories} & \textbf{Num.} & \textbf{Sub Category} & \textbf{Num.} \\
\midrule
\multirow{4}{*}{Content Presentation} & \multirow{4}{*}{28} 
 & Static Page Generation   & 20 \\
 & & Dynamic Content Rendering & 18 \\
 & & Data Visualization       & 36 \\
 & & Media Display            & 6 \\
\midrule
\multirow{5}{*}{User Interaction} & \multirow{5}{*}{49}
 & Form Systems      & 40 \\
 & & Authentication   & 18 \\
 & & Real-time Features & 20 \\
 & & E-commerce       & 22 \\
 & & AI Integration   & 19 \\
\midrule
\multirow{4}{*}{Data Management} & \multirow{4}{*}{24}
 & CRUD Operations   & 29 \\
 & & API Integration  & 20 \\
 & & Big Data         & 12 \\
 & & File Handling    & 5 \\
\midrule
\textbf{Total} & \textbf{101} & & \\
\bottomrule
\end{tabular}
}
\end{table}

\subsubsection{Functionality Metrics}
We adopt a UI agent-based evaluation for functionality correctness. We utilize BrowserUse, one of the SOTA web navigation UI agents\footnote{https://browser-use.com/posts/sota-technical-report}, to execute test operations and verify outcomes. Each test case’s operation and expected result are encoded into a standardized prompt, which directs the agent to simulate interactions, analyze trajectories and screenshots, and return \textit{YES}, \textit{NO}, or \textit{PARTIAL} for requirement fulfillment. When the interaction limit is reached without completion, a decision prompt induces the agent to make a final judgment. We employ Claude-4-Sonnet as the agent’s engine. All the instructions for the evaluation agent are adopted from~\cite{lu2025webgen}.

\subsubsection{Visual Metrics}
Apart from functionality correctness, we also evaluate the visual quality of generated websites. We use a set of metrics covering rendering success, content relevance, layout harmony, design modernness, and similarity with the design image. These metrics are formulated into a prompt and assessed by Claude-4-Sonnet, which assigns a score from 1 to 5 (higher is better) to quantify the overall aesthetics and visual consistency of each website. The procedure and prompts for visual quality evaluation are adopted from~\cite{lu2025webgen}. The prompt for deciding visual similarity is shown below:

\begin{tcolorbox}
[colback=white,colframe=black!50,title=Visual Similarity Evaluation Prompt]
\footnotesize
You are an expert web designer and evaluator. Your task is to assess how well a generated website matches a given design image. Consider the following aspects:  1. \textbf{Layout:} Does the overall structure of the website (e.g., positioning of sections, grids, spacing) align with the design image?  
2. \textbf{Components:} Are the required UI elements (e.g., buttons, forms, images, navigation bars) present and placed correctly?  
3. \textbf{Color Scheme:} Are the background, text, and component colors consistent with the design image?  
4. \textbf{Size and Proportion:} Do elements (e.g., images, text blocks, buttons) have similar relative sizes and proportions compared to the design image?  
5. \textbf{Visual Fidelity:} Does the overall look and feel of the website closely resemble the design image?  

\textbf{Instruction:}  
Compare the website rendering against the design image. Provide an overall similarity score from \textbf{1 to 5} (1 = very poor match, 5 = almost identical).  
\end{tcolorbox}

\section{Result \& Analysis}
\subsection{RQ1: Effectiveness of \methodname}
\subsubsection{Overall Effectiveness}
This section evaluates the overall performance of \methodname compared with the baselines. The result is presented in Table~\ref{tab:overall-results}. Accuracy is computed using the formula $\text{Accuracy} = \frac{N_{\text{Yes}} + 0.5 \times N_{\text{Partial}}}{N_{\text{Total}}} \times 100\%,$ where $N_{\text{Yes}}$ and $N_{\text{Partial}}$ denote the number of test cases assessed as \textsc{Yes} and \textsc{Partial}, respectively, and $N_{\text{Total}}$ is the total number of test cases.

We can see that \methodname consistently outperforms both Bolt.diy and Cursor across most metrics. With GPT-4.1, \methodname achieves the highest accuracy (78.2\%), a relative improvement of +30\% over Cursor (60.2\%) and more than triple the performance of Bolt.diy (25.6\%). Similarly, with Claude-4-Sonnet, \methodname attains 70.2\%, surpassing Cursor (63.8\%) and substantially outperforming Bolt.diy (44.5\%). Importantly, \methodname has the lowest \textsc{No} rate (9.5\% with GPT-4.1 and 15.0\% with Claude-4-Sonnet), showing that it is less likely to fail functional requirements outright.

Reliability is another advantage: \methodname shows no failures to start in either configuration, while both baselines exhibit frequent launch failures (e.g., 80.0\% for Bolt.diy with GPT-4.1 and 40.0\% for Cursor with GPT-4.1). This highlights \methodname’s robustness in generating runnable applications without manual intervention.

On user-facing criteria, \methodname also delivers competitive results. Its appearance scores are consistently high (4.0 with GPT-4.1), and with Claude-4-Sonnet it achieves the best visual similarity (4.3), indicating faithful reproduction of UI designs. Although the similarity score is lower with GPT-4.1 (2.3), this suggests that the choice of model can affect the trade-off between functional accuracy and visual fidelity.

\begin{table}[t]
\centering
\caption{Evaluation of three code-agent frameworks using different proprietary and open-source models. Accuracy is computed using a weighted score, where YES samples are weighted by 1 and PARTIAL samples are weighted by 0.5; the total score is then divided by the number of test cases. The highest accuracy and appearance scores are marked in \textbf{bold}. Appearance score and Vis. Similarity score are between 1-5, other metrics are in percentage.}
\label{tab:overall-results}
\resizebox{\linewidth}{!}{
\begin{tabular}{lccccccc}
\toprule
\textbf{Test Name} & Yes Rate & Partial Rate & No Rate & Fail to Start & \textbf{Accuracy} & \textbf{Appearance}  & \textbf{Vis. Similarity}\\
\midrule
\multicolumn{8}{l}{\textbf{\methodname}} \\

GPT-4.1          & \textbf{65.9} & 24.6 & \textbf{9.5} & \textbf{0.0} & \textbf{78.2} & \textbf{4.0} & 2.3 \\
Claude-4-Sonnet  & 55.0 & 30.0 & 15.0 & \textbf{0.0} & 70.2 & 3.7 & \textbf{4.3} \\

\midrule
\multicolumn{8}{l}{\textbf{Bolt.diy}} \\

GPT-4.1          & 23.1 & 7.7 & 69.2 & 80.0 & 25.6 & \textbf{4.0} & 3.0 \\
Claude-4-Sonnet  & 26.9 & 34.6 & 38.5 & 20.0 & 44.5 & 3.4 & 2.6 \\

\midrule
\multicolumn{8}{l}{\textbf{Cursor}} \\

GPT-4.1          & 44.7 & 26.3 & 28.9 & 40.0 & 60.2 & 3.3 & 3.0 \\
Claude-4-Sonnet  & 50.0 & 25.0 & 25.0 & 20.0 & 63.8 & 3.8 & 3.3 \\

\bottomrule
\end{tabular}

}
\end{table}

\subsubsection{Detailed Result}
Besides overall results, we calculate the accuracy for each category of instructions and test cases in Table~\ref{tab:category-results}. The result shows that \methodname consistently achieves the strongest or near-strongest performance across most categories. For instruction categories, \methodname with GPT-4.1 excels in handling user interaction (91.7\%) and data management (78.6\%), outperforming both Bolt.diy and Cursor. Claude-4-Sonnet achieves the best performance in content presentation (71.4\%), while maintaining competitive scores in other categories. For test case categories, \methodname also demonstrates superiority. With GPT-4.1, it achieves the highest accuracy in functionality (66.7\%) and data display (87.5\%), while Claude-4-Sonnet reaches perfect performance in design validation (100\%). 

\begin{table}[t]
\centering
\caption{Category-wise evaluation results. The first three columns represent categories of website-generation instructions, while the last three represent categories of test cases. All metrics are in percentage. N.A. implies all generated applications failed to start in the category.}
\label{tab:category-results}
\resizebox{\linewidth}{!}{
\begin{tabular}{lcccccc}
\toprule
\textbf{Test Name} & 
\multicolumn{3}{c}{\textbf{Instruction Categories}} & 
\multicolumn{3}{c}{\textbf{Test Case Categories}} \\
\cmidrule(lr){2-4} \cmidrule(lr){5-7}
& \textbf{Content} & \textbf{User Interact.} & \textbf{Data Manage.} 
& \textbf{Functionality} & \textbf{Data Display} & \textbf{Design Validation} \\
\midrule
\multicolumn{7}{l}{\textbf{\methodname}} \\
GPT-4.1          & 64.3 & \textbf{91.7} & \textbf{78.6} & \textbf{66.7} & \textbf{87.5} & 83.3 \\
Claude-4-Sonnet  & \textbf{71.4} & 75.0 & 64.3 & 61.1 & 68.8 & \textbf{100.0} \\
\midrule
\multicolumn{7}{l}{\textbf{Bolt.diy}} \\
GPT-4.1          & 42.9 & 8.3 & N.A. & 8.3 & 40.0 & 50.0 \\
Claude-4-Sonnet  & 47.0 & 36.9 & 57.1 & 28.3 & 55.0 & 61.1 \\
\midrule
\multicolumn{7}{l}{\textbf{Cursor}} \\
GPT-4.1          & 61.9 & 48.6 & \textbf{78.6} & 55.0 & 59.1 & 58.3 \\
Claude-4-Sonnet  & 65.5 & 60.6 & 71.4 & 55.0 & 67.4 & 66.7 \\
\bottomrule
\end{tabular}

}
\end{table}

\begin{tcolorbox}
[colback=gray!30,colframe=gray!30]
Answer to RQ1: \methodname is effective in automating full-stack web development. It achieves higher accuracy, stronger reliability, and more consistent visual quality than baseline frameworks across both overall and category-wise evaluations and across two different SOTA LLM backbones.
\end{tcolorbox}

\subsubsection{RQ2: Ablation Study on Test Case Generation Agent}

\begin{table}[t]
\centering
\caption{Comparison of \methodname with straightforward test generation versus proposed test generation. Accuracy is reported as percentages. The best performance in each category is marked in \textbf{bold}.}
\label{tab:testgen-comparison}
\resizebox{0.7\linewidth}{!}{
\begin{tabular}{lcc}
\toprule
\textbf{Setting} & \textbf{Strtfwd. Test Gen + \methodname} & \textbf{Multi-Step Test Gen + \methodname} \\
\midrule
Functionality        & 33.3  & \textbf{61.1} \\
Data Display         & \textbf{75.0} & 68.8 \\
Design Validation    & \textbf{100.0} & \textbf{100.0} \\
\midrule
\textbf{Overall Acc.}      & 59.1  & \textbf{70.2} \\
\textbf{Vis. Similarity (1-5)}   & \textbf{4.3}  & \textbf{4.3} \\
\bottomrule
\end{tabular}

}
\end{table}

In \methodname, we design a multi-step test case generation agent that consists of three stages: high-level requirement extraction, detailed requirement completion, and soap opera test generation. This design aims to progressively decompose user instructions into testable requirements while ensuring coverage of both front-end and back-end aspects. To assess the necessity of such a design, we compared \methodname with the multi-step agent against a variant using a straightforward test case generation agent. In the latter, the model directly produces test cases from the initial user input using the following prompt:

\begin{tcolorbox}
[colback=white,colframe=black!50,title=Straightforward Test Case Generation Prompt]
\footnotesize
You are an expert AI Product Manager. Your task is to analyze the following user-provided instruction for webpage development and break it down into a detailed list of requirements and test cases.

Your analysis must: 1) Capture every detail from the user’s instructions without omission, no matter how minor. 2) Go beyond front-end representation by inferring and specifying necessary backend components (e.g., databases, APIs, integrations) to ensure the feature is fully functional and realistic. 3) Include details on database setup, API calls, and integration requirements wherever applicable. 4) Prefer real integrations over placeholders; when exact details are unavailable, clearly specify a placeholder rather than inventing information.

The final output should be a JSON array of this format: [JSON Format Specification]

\end{tcolorbox}

The comparison in Table~\ref{tab:testgen-comparison} reveals several insights. First, the multi-step agent yields a substantially higher functionality accuracy (61.1\% vs.\ 33.3\%), showing that progressive refinement better captures system-level requirements and functional dependencies. While the straightforward approach achieves slightly higher accuracy in data display (75.0\% vs.\ 68.8\%), this gain is offset by its weakness in functionality, suggesting that it focuses more on surface-level outputs rather than end-to-end correctness. Both approaches achieve perfect performance in design validation (100.0\%), reflecting that visual styling requirements are easier to capture regardless of test case generation method.

Our further inspection shows that the performance gain in functionality is two-fold: (1) the high-level requirement list guides the development agent toward a deeper understanding of the task and more comprehensive planning, thereby improving code generation quality; and (2) the detailed test cases derived from refined requirements enable the testing agent to provide more targeted and fine-grained feedback, which helps the development agent refine implementations more effectively. These results demonstrate that decomposing test case generation into multiple steps is more effective than a single-shot approach, as it leads to stronger functional reliability and balanced coverage across categories while maintaining high design fidelity.

\begin{tcolorbox}
[colback=gray!30,colframe=gray!30]
Answer to RQ2: The multi-step design of the test case generation agent is effective. It improves functionality accuracy and overall reliability by guiding the development agent with structured requirements and enabling the testing agent to deliver more targeted feedback.
\end{tcolorbox}

\subsection{RQ3: Ablation Study on Testing Agent Feedback}
To study the effectiveness of the testing agent feedback, we compare the performance of \methodname without feedback and with different rounds of iterative feedback. In the no-feedback setting, the testing agent is disabled and only the initial output of the development agent is evaluated. In the feedback settings, we vary the number of feedback iterations by adjusting $max\_iter = 1, 2, 3$ to examine how additional refinement cycles impact performance.

The results in Table~\ref{tab:feedback-rounds} indicate several key trends. First, introducing 1-2 rounds of feedback does not immediately improve functional correctness or overall test accuracy—in fact, accuracy drops to 25.3\%, suggesting that 1-2 rounds may be insufficient for the model to meaningfully refine the code (e.g., the agent may introduce new bugs when refining the code). However, as the number of rounds increases, improvements emerge. With three rounds of feedback, \methodname achieves the best performance across nearly all metrics: functionality rises to 61.1\%, data display to 68.8\%, and test accuracy to 70.2\%, substantially outperforming the no-feedback baseline (58.3\%).

Visual quality also benefits from iterative refinement. The visual similarity score increases from 2.3 (no feedback) to 4.3 with three rounds, indicating that feedback not only strengthens correctness but also helps align the generated application more closely with design specifications. Importantly, design validation remains perfect (100.0\%) across all settings, showing that feedback mainly influences deeper functional and data-related aspects rather than surface-level styling.

\begin{tcolorbox}
[colback=gray!30,colframe=gray!30]
Answer to RQ3: Testing agent feedback is effective in enhancing the quality of generated applications. Multiple rounds of feedback enable the agent to iteratively fix bugs and improve functionality, accuracy, and visual fidelity.
\end{tcolorbox}

\begin{table}[t]
\centering
\caption{Effectiveness of the testing agent in \methodname. We compare \methodname without feedback and with different rounds of test feedback ($max\_iter=1,2,3$). TDD Pass Rate is the pass rate of generated test cases in the TDD workflow. Test Acc. is the overall accuracy of WebGen-Bench test cases.}
\label{tab:feedback-rounds}
\resizebox{0.7\linewidth}{!}{
\begin{tabular}{lcccc}
\toprule
\textbf{Evaluation Setting} & \textbf{No Feedback} & \textbf{1 Round} & \textbf{2 Rounds} & \textbf{3 Rounds} \\
\midrule
Functionality     & 44.4 & 33.3 & 33.3 & 61.1 \\
Data Display      & 56.3 & 56.3 & 43.8 & 68.8 \\
Design Validation & 100.0 & 100.0 & 100.0 & 100.0 \\
\midrule
\textbf{TDD Pass Rate} & 19.6 & 16.3 & 30.4 & 33.8 \\
\textbf{Test Acc.} & 58.3 & 25.3 & 25.2 & 70.2 \\
\textbf{Vis. Similarity (1-5)} & 2.3 & 3.7 & 3.3 & 4.3 \\
\bottomrule
\end{tabular}

}
\end{table}

\subsection{RQ4: Advantages for Developers Over Existing Open-Source Tools}
We assess the practicality of \methodname in accelerating the development workflow. Specifically, we implement \methodname into a user-friendly demo tool and recruit three developers (two research staff who have previously developed at least two web applications, and one front-end developer from a startup company) to participate in our study, following the methodology of Chen et al.~\cite{Chen2018FromUI}. Each participant is tasked with developing a web application from a requirement in WebGen-Bench. Participants are asked first to develop the application using \methodname and then to develop the same application again with an open-source industry framework, Bolt.diy. In both cases, they refine the application until it reaches a satisfactory state.

We record the manual intervention time (inputting prompts, testing, and revising code) for each participant, as shown in Table~\ref{tab:manual-vs-accuracy}, and capture their development processes through screen recordings.

The results highlight a key difference between the two methods. With Bolt.diy, participants spent an average of 4.7 minutes on manual interventions across a total of 15.2 minutes, requiring three rounds of interaction and 74 additional words of prompting. Although 4.7 minutes of manual effort may appear modest, this time is spread across the entire 15.2-minute session. For example, a participant first inputs the initial prompt, then after 3 minutes of agent execution must test and provide feedback, and again after another refinement cycle, more feedback is required. As a result, the agent demands the user’s continuous attention, preventing them from focusing on other tasks.

By contrast, with \methodname, participants required no manual interventions. Once the initial requirement was provided, the system autonomously handled development, testing, and refinement. Although the total development time was slightly longer (18.7 minutes), users had the full duration free to focus on other work without interruption.

This finding underscores a central advantage of \methodname: it shifts the workload from continuous prompt engineering and manual testing to autonomous, feedback-driven refinement. Developers are freed from micromanaging iterations and can concentrate on higher-level design decisions or parallel tasks, leading to a more efficient and less disruptive workflow.

\begin{tcolorbox}
[colback=gray!30,colframe=gray!30]
Answer to RQ4: \methodname provides advantages over existing open-source tools by eliminating the need for manual interventions. Existing open-source tools require scattered user input and constant attention, while \methodname allows developers to remain fully disengaged during generation, reducing workload and enabling them to focus on other tasks while the system completes development autonomously.
\end{tcolorbox}

\begin{table}[t]
\small
\centering
\caption{Comparison of manual intervention time.}
\label{tab:manual-vs-accuracy}
\begin{tabular}{lcccc}
\toprule
\textbf{Method} & \textbf{Manual/Total Time (min)} & \textbf{Intervention Frequency}  & \textbf{Additional Prompt Len. (word)}  \\
\midrule
Bolt.diy  & 4.7/15.2 & 3.0 & 74.0\\
\methodname &  0.0/18.7 & 0.0 & 0.0\\

\bottomrule
\end{tabular}
\end{table}

\section{Discussion}
\label{sec:discussion}
\subsection{Accuracy of UI Agent Testing}
Our evaluation relies on UI agent testing; to evaluate the accuracy of the UI agent testing process, we randomly sampled 5 generated web applications and manually verified the testing results produced by the agent, covering 28 test cases in total. We focus on Claude-4-Sonnet, given its high accuracy, and compare two UI agents: BrowserUse (used in our work) and WebVoyager~\cite{He2024WebVoyagerBA} (used in WebGen-Bench). Manual testing serves as ground truth and requires precision; thus, two annotators independently labeled each case, with disagreements resolved by a third annotator. The Alignment Rate is defined as $\text{Alignment Rate} = \frac{N_{\text{Manual}=\text{Agent}}}{N_{\text{Total}}} \times 100\%,$ where $N_{\text{Manual}=\text{Agent}}$ denotes the number of test cases where the agent-generated result matches the manual annotation. Table~\ref{tab:manual-testing} reports performance measured by manual inspection and agent testing, along with alignment rates.

From these results, we observe: \textbf{(1) the Yes/No rates and Accuracy reported by BrowserUse closely track manual testing}, with only minor fluctuations of 0.8–6.2\%; \textbf{(2) the overall alignment rate is satisfactory}, with 23 out of 28 cases perfectly aligned, yielding an 82.8\% alignment rate.

To further interpret the relatively high Partial rate and the fluctuations in Yes/No/Accuracy, we examined the misaligned cases. We found that all agent judgment errors originate from \textit{partially correct} cases, which often involve subjective strictness (e.g., whether an error should be raised when invalid input is detected). The agent tended to map human-annotated Yes or No cases into Partial, and occasionally labeled human-perceived Partial cases as Yes or No. \textbf{When restricting to strictly Yes/No cases, the alignment rate improves to 100\% (23 out of 23).}

\begin{table}[t]
\small
\centering
\caption{Performance (\%) reported by different testing entities and alignment rate (\%) between UI agent testing results and manual testing results. Numbers in brackets indicate the difference from manual testing results.}
\label{tab:manual-testing}
\begin{tabular}{lcccccc}
\toprule
\textbf{Testing Method} & Yes Rate & Partial Rate & No Rate & \textbf{Acc} & \textbf{Align.}  & \textbf{Yes/No Align.}\\
\midrule
WebVoyager
& 42.9 {\scriptsize (-23.8)} 
& 25.0 {\scriptsize (+11.7)} 
& 32.1 {\scriptsize (+12.1)} 
& 55.4 {\scriptsize (-18.0)} 
& 46.4 
& 66.7  \\

BrowserUse (\textbf{\methodname}) 
& 62.1 {\scriptsize (-4.6)} 
& 24.1 {\scriptsize (+10.8)} 
& 13.8 {\scriptsize (-6.2)} 
& 74.1 {\scriptsize (+0.8)} 
& 82.8 
& 100.0 \\
\midrule
Manual                 
& 66.7 & 13.3 & 20.0 & 73.3 & -- & --\\
\bottomrule
\end{tabular}
\end{table}

\subsection{Failure Analysis and Future Works}

\subsubsection{Scaling Feedback Rounds}
During the experiment in RQ3, we observed that increasing the number of feedback rounds beyond five introduces greater variance in overall accuracy. On the one hand, additional rounds give the agent more opportunities to refine code and resolve existing bugs. On the other hand, as the codebase becomes increasingly complex after successive edits, it becomes more susceptible to the introduction of new errors. Consequently, excessive feedback cycles may not consistently yield performance gains.

This variation can be mitigated by straightforward selection strategies, such as executing multiple rounds (e.g., ten rounds) and selecting the best-performing output. In future work, we plan to explore more efficient mechanisms, such as branching and rollback strategies, that can scale feedback rounds while maintaining stability.

\subsubsection{Open-Source LLMs}
Beyond the widely adopted proprietary models Claude-4-Sonnet and GPT-4.1, we also conducted preliminary experiments with two popular open-source models: DeepSeek-V3.1, known for strong coding ability, and Qwen-2.5-VL-72B, known for strong visual capacity. Although \methodname is adaptable to these models, we identified several challenges:

\begin{itemize}
\item \textbf{Output Formatting:} \methodname requires strict output formats. The Test Generation Agent must produce JSON outputs, while the Development Agent depends on XML-structured outputs for parsing context and executing edits. Both DeepSeek-V3.1 and Qwen-2.5-VL-72B frequently failed these formatting constraints, leading to parsing errors or truncated outputs, which in turn caused workflow failures.
\item \textbf{Coding Ability:} Despite Qwen-2.5-VL-72B’s strong visual reasoning, its limited coding ability resulted in a high number of fail-to-start cases.
\item \textbf{Visual Ability:} DeepSeek-V3.1, while strong in code generation, lacks visual reasoning capacity and thus cannot independently power the visual-driven testing agent. When paired with Claude-4-Sonnet to provide testing feedback, performance remained limited due to formatting inconsistencies.
\end{itemize}

These observations suggest that current open-source LLMs face difficulties in agent workflows that demand structured output, code generation, and multimodal reasoning. As part of future work, we aim to identify or train models with stronger agentic capacities, particularly in enforcing output formats and balancing coding with visual reasoning.

\section{Threat to Validity}
We have identified the following threats to the validity of our work:
\paragraph{Effectiveness of the Agentic System Design}
\methodname is built upon an intricate agentic workflow, and its performance may be constrained by the effectiveness of individual components. To mitigate this concern, we conducted an extensive ablation study, which validates the contributions of the test case generation module, the testing module, and the overall TDD pipeline. These results provide evidence that the system design is both robust and effective.

\paragraph{Reliability of UI Agent Testing}
Our evaluation relies on UI agent testing, which requires high levels of reliability and accuracy. To ensure trustworthiness, we supplemented automated testing with manual inspection. The results show that the performance reported by the UI agent closely aligns with human judgments, with an overall alignment rate that we consider satisfactory.

\section{Conclusion}
In summary, this work presents \methodname, the first TDD-enabled LLM-agent framework capable of generating end-to-end full-stack web applications from natural language or visual specifications. By integrating automated requirement extraction, test case generation, iterative code refinement, and user interaction simulation, \methodname overcomes the limitations of existing approaches that focus solely on front-end generation. Our experiments demonstrate that \methodname not only ensures both functional correctness and visual fidelity but also significantly reduces failure rates compared to state-of-the-art baselines. These results highlight the potential of TDD-driven LLM agents to advance the automation of full-stack development, paving the way for more reliable, efficient, and scalable web application generation.

\section*{Data Availability}
The code for \methodname is available at \url{https://github.com/yxwan123/TDDev} for replication and future research.

\bibliographystyle{ACM-Reference-Format}
\bibliography{reference}


\end{document}